\documentclass{ifacconf}

\usepackage{graphicx}          % Include this line if your 

 \usepackage{amsfonts}
\usepackage{amsmath}
\usepackage{amssymb}
\usepackage{epsfig}
\usepackage{algorithm}
\usepackage{algorithmicx}
\usepackage{algpseudocode}
\usepackage{color}
\usepackage{subfig}

\newtheorem{remark}{Remark}
\newtheorem{proposition}{Proposition}

\newtheorem{corollary}{Corollary}
\newtheorem{definition}{Definition}
\newtheorem{theorem}{Theorem}
\newtheorem{problem}{Problem}

\newcommand{\FNab}{{\cal F}(a,b)}
\newcommand{\RNset}{R_N(a^{\mathbb{X}},b^{\mathbb{X}})}
\newcommand{\VNset}{\mathcal{V}(a,b)}
\newcommand{\RNab}{R_N(a,b)}

\newcommand{\TS}{{\cal T}}
\newcommand{\XO}{\ensuremath{\mathbb{X}_0}}
\newcommand{\AUTOMATON}{\ensuremath{\mathcal{A}}}
\newcommand{\Fset}{{\cal E}}

\newcommand{\minSAP}{\ensuremath{\min_{v \in \mathcal{V}(F)}  ((h(v) + Bk(v))_i) }}
\newcommand{\minSBP}{\ensuremath{  \min_{v \in \mathcal{V}( {\overline{F}})}  ((h(v) + Bk(v))_i)}}

\newcommand{\LTLUNTIL}{\ensuremath{\mathcal{U}}}

\newcommand{\LTLEVENTUALLY}{\ensuremath{\mathcal{F}}}
\newcommand{\PREDSET}{\ensuremath{\Pi}}
\newcommand{\TIMEBOUND}{\ensuremath{T}}

\newcommand{\ARUN}{\ensuremath{r}}
\newcommand{\scheck}{\textit{scheck2\ }}

\begin{document}

\begin{frontmatter}

\title{Time-Constrained Temporal Logic Control of Multi-Affine Systems}
\thanks[footnoteinfo]{This work was partially supported at Boston University by grants AFOSR YIP FA9550-09-1-0209, ARO W911NF-09-1-0088, NSF CNS-0834260, ONR 
MURI N00014-10-10952, and ONR MURI N00014-09-1051.}

\author[First]{Ebru Aydin Gol}
\author[First]{Calin Belta}

\address[First]{Boston University, Boston, MA 02215, USA \\e-mail: \{ebru,cbelta\}@bu.edu}

\begin{abstract}
In this paper, we consider the problem of controlling a dynamical system such that its trajectories satisfy a temporal logic property in a given amount of time. We focus on multi-affine systems and specifications given as syntactically co-safe linear temporal logic formulas over rectangular regions in the state space. The proposed algorithm is based on the estimation of time bounds for facet reachability problems and solving a time optimal reachability problem on the product between a weighted transition system and an automaton that enforces the satisfaction of the specification. A random optimization algorithm is used to iteratively improve the solution. 
\end{abstract}
\end{frontmatter}

\section{Introduction}\label{sec:intro}
Temporal logics and model checking algorithms have been primarily used for specifying and verifying correctness of software and hardware systems. 
Due to their expressivity and resemblance to natural language, temporal logics have gained popularity as specification languages in other areas including dynamical systems. Recently, there has been increasing interest in formal synthesis of dynamical systems, where the goal is to generate a control strategy for a dynamical system from a specification given as a temporal logic formula, such as Linear Temporal Logic (LTL)  (\cite{Kloetzer:2008,TP03,Girard:2010}), or fragments of LTL, such as  
GR(1) (\cite{Hadas-ICRA07,Tok-Ufuk-Murray-CDC09}) and syntactically co-safe LTL (\cite{Kavraki:MPlanning}).

We focus on a particular class of nonlinear affine
control systems, where the drift
is a multi-affine vector field ({\em i.e.,} affine in each state
component), the control distribution is constant, and the
control is constrained to a convex set. This class of
dynamics includes the
Euler, Volterra (\cite{Volterra1926}) and Lotka-Volterra
(\cite{Lotka1925}) equations, attitude and velocity control systems
for aircraft (\cite{vanderSchaft}) and underwater vehicles
(\cite{Belta-ICRA2004}), and models
of biochemical networks (\cite{jong2002}). 
In \cite{Belta-TAC06}, the authors studied the problem of synthesizing a state feedback controller such that the trajectories originating in a rectangle leave it through a specified facet. These results were generalized in \cite{Habets2006} by allowing the trajectories to leave through a set of exit facets. 

In this paper, we consider the following problem: given a multi-affine control system and a syntactically co-safe LTL formula over rectangular subregions of the state space, find a set of initial states for which there exists a control strategy such that all the trajectories of the closed-loop system satisfy the formula within a given time bound. Syntactically co-safe LTL formulas can be used to describe finite horizon specifications such as target reachability with obstacle avoidance: ``always avoid obstacle $O$ until reaching target $T$", sequencing constraints ``do not go to $A$ or $B$ unless $C$ was visited before", and more complex temporal and Boolean logic combinations of these. Our approach to this problem consists of two main steps. First, we construct a finite abstraction of the system by solving facet reachability problems on a rectangular partition of the state space. We build on the results from \cite{Belta-TAC06,Habets2006} to derive bounds for the exit times of the trajectories. 
Second, we solve time optimal reachability problems on the product between the abstraction and an automaton that enforces the satisfaction of the specification. 
We propose an iterative refinement procedure via a random optimization algorithm.

Finite abstractions for controlling dynamical systems have been widely used, e.g by \cite{TP03}. 
Time optimal control of dynamical systems through abstractions has been studied by \cite{Mazo:2011} and \cite{Girard:2010Opt}. In both cases, an optimal controller is synthesized for an approximate abstraction, which is then mapped to a suboptimal solution for the original system for specifications given in the form of ``reach and avoid'' sets. While our solution also involves an optimal control problem on the abstraction, our automata-theoretic approach allows for richer, temporal logic 
control specifications.

The remainder of the paper is organized as follows. We review some notions necessary throughout the paper in Sec.~\ref{sec:pre} before formulating the problem and outlining the approach in Sec.~\ref{sec:prob}. 
A review of facet reachability problems and the derivation of the exit time bounds are presented in Sec.~\ref{sec:control}. 
The control strategy providing a solution to the main problem is described in Sec.\ref{sec:controlStrategy} and the random optimization method for refinement is given in Sec.~\ref{sec:optimization}. 
An example is given in Sec.~\ref{sec:casestudy} and conclusions are summarized in Sec.~\ref{sec:conclusion}.

%%%%%%%%%%%%%%%%%%%%%%%%%%%%%%%%%%%%%%%%%%%

\section{Preliminaries}\label{sec:pre}

\subsection{Transition systems and linear temporal logic}
\begin{definition}
A weighted transition system is a tuple $\TS=(Q, \Sigma, \delta, O, o, w)$, where $Q$ and $\Sigma$ are sets of states and inputs, $\delta: Q \times \Sigma \longrightarrow 2^Q$ is a transition map, $O$ is a set of observations, $o:Q \longrightarrow O$ is an observation map, and $w : Q \times \Sigma \longrightarrow \Rset_+$ is a map that assigns a positive weight to each state and input pair.
\end{definition}

$\delta(q,\sigma)$ denotes the set of successor states of $q$ under the input $\sigma$. If the cardinality of $\delta(q,\sigma)$ is one, the transition $\delta(q,\sigma)$ is deterministic. A transition system $\TS$ is called deterministic if all its transitions are deterministic.

A finite input word $\sigma_1 \ldots \sigma_n$, $\sigma_i\in\Sigma$, $i=1,\ldots,n$ and an initial state $q_0 \in Q$ define a trajectory $r=q_0 \ldots q_n$ of the system with the property that $q_{i+1} \in \delta(q_i, \sigma_{i+1})$ for all $0 \leq i \leq n-1$. The cost $J^{\TS}(r)$ of trajectory $r$ is defined as the sum of the corresponding weights, i.e., 
\[
J^{\TS}(r) = \sum_{i=0}^{n-1} w(q_i, \sigma_{i+1}).
\]
A trajectory $r=q_0\ldots q_n$ produces a word $o(q_0)\ldots$ $o(q_n)$.

\begin{definition}( \cite{Vardi:safety}) A syntactically co-safe LTL (scLTL) formula over a set of atomic propositions $\PREDSET$ is inductively defined as follows:
\begin{equation}
	\Phi := \pi | \neg \pi | \Phi \vee \Phi |  \Phi \wedge \Phi |  \Phi \LTLUNTIL \Phi | \LTLEVENTUALLY \Phi,
\end{equation}

where $\pi\in\Pi$ is an atomic proposition, $\neg$ (negation), $\vee$ (disjunction), $\wedge$ (conjunction) are Boolean operators, and $\LTLUNTIL$ (``until''), and $\LTLEVENTUALLY$ (``eventually'') are temporal operators \footnote{The scLTL syntax usually includes a ``next" temporal operator. We do not use it here 
because it is irrelevant for the particular semantics of continuous trajectories that we define later.}. 
\end{definition}

The semantics of scLTL formulas is defined over infinite words over $2^\PREDSET$. Informally, 
$\pi_1\LTLUNTIL \pi_2$ states that $\pi_1$ is true until $\pi_2$ is true and $\pi_2$ becomes eventually true in a word; $\LTLEVENTUALLY \pi_1$ states that $\pi_1$ becomes true at some position in the word. More complex specifications can be defined by combing temporal and Boolean operators (see Eqn. (\ref{eq:spec_case_study})). 

An important property of scLTL formulas is that, even though they have infinite-time semantics,
their satisfaction is guaranteed in finite time. Explicitly, for any scLTL formula $\Phi$ over $\PREDSET$,
any satisfying infinite word over $2^\PREDSET$ contains a satisfying finite prefix. 

\begin{definition} A deterministic finite state automaton (FSA) is a tuple $\AUTOMATON = (S, \Pi, \delta_\AUTOMATON, S_0, F)$ where $S$ is a finite set of states, $\Pi$ is an input alphabet, $S_0 \subseteq S$ is a set of initial states, $F \subseteq S$ is a set of final states, and $\delta_\AUTOMATON : S \times \Pi \longrightarrow S$ is a deterministic transition relation.
\end{definition}

An accepting run $\ARUN_\AUTOMATON$ of an automaton $\AUTOMATON$ on a finite word $w=w_0\ldots w_d$ over $\Sigma$ is a sequence of states $\ARUN_\AUTOMATON=s_0\ldots s_{d+1}$ such that $s_0 \in S_0$, $s_{d+1} \in F$ and $\delta_\AUTOMATON(s_i,w_i) = s_{i+1}$ for all $i=0,\ldots,d$. 
For any scLTL $\Phi$ formula over $\PREDSET$, there exists a FSA $\AUTOMATON$ with input alphabet $2^\PREDSET$ that accepts the prefixes of all the satisfying words. 
There are algorithmic procedures and off-the-shelf tools, such as \scheck by \cite{Latvala:scheck},
for the construction of such an automaton.

\begin{definition}\label{def:product_automaton}
Given a weighted transition system $\TS=(Q, \Sigma, \delta, O, o, w)$ and a FSA $\AUTOMATON = (S, \PREDSET, \delta_\AUTOMATON, S_0, F)$ with $O=\PREDSET$, their product automaton is a FSA $\AUTOMATON^P = (S_P, \Sigma, \delta_P, S_{P0}, F_P)$ where $S_P=Q\times S$ is the set of states, $\Sigma$ is the input alphabet, 
$\delta_P: S_P \times \Sigma \longrightarrow 2^{S_P}$ is the transition relation with $\delta_P((q,s), \sigma)=\{(q',s') \mid q' \in \delta(q, \sigma), \delta_\AUTOMATON(s, o(q)) = s'\}$,
$S_{P0}=Q\times S_0$ is the set of initial states, and $F_P=Q\times F$ is the set of final states.

An accepting run $r_P = (q_0,s_0)\ldots (q_n,s_n)$ of $\AUTOMATON^P$ defines an accepting run $s_0\ldots s_n$ of $\AUTOMATON$ over input word $o(q_0)\ldots o(q_{n-1})$.
The weight function of the transition system can directly be used to assign weights to transitions of $\AUTOMATON^P$, i.e., we can define a weight function for the product automaton in the form $w_P( \delta_P((q,s), \sigma)) = w( \delta(q,\sigma))$. 
The corresponding cost for a run $r_P = (q_0,s_0)\ldots (q_n,s_n)$ of $\AUTOMATON^P$ over $\sigma_1\ldots \sigma_n$ is defined as 
\[J^P(r_P)= \sum_{i=1}^n w_P(\delta_P((q_{i-1},s_{i-1}), \sigma_i)).\] 
\end{definition}
\newcommand{\SIGNLETTER}{\gamma}
\subsection{Rectangles and multi-affine functions}
For $N \in \Nset$, an $N$-dimensional rectangle $\RNab \subset \Rset^N$ is
characterized by two vectors $a = (a_1,\ldots,a_N)$ and $b =
(b_1,\ldots,b_N)$ with the property that $a_i < b_i$ for all $i =
1,\ldots,N$:
\begin{equation}\label{eq:rec}
\begin{array}{l}
  \RNab =  \{x \in \Rset^N \mid \forall i \in \{1,\ldots,N\}:\; a_i \leq x_i \leq b_i\}.
\end{array}
\end{equation}

Let $\VNset$ and $\FNab$ be the set of vertices and facets of 
of $\RNab$, respectively. 
Let $F^{\pm e_i}$ denote the facet with normal $\pm e_i$, where 
$e_i$, $i=1,\ldots,N$ denote the standard basis of $\Rset^N$.
For a facet $F\in \FNab$, ${\cal V}(F)$  denotes its set of vertices
and $n_F$ denotes its outer normal. For a vertex $v \in \VNset$, ${\cal F}_v$ denotes the set of facets containing $v$. 

\begin{definition}
  \label{def:multiaffine}
  A {\em multi-affine function} $h: \Rset^N \longrightarrow
  \Rset^q$ (with $N,q \in \Nset$) is a function that is affine in
  each of its variables, i.e., $h$ is of the form
  \[h(x_1,\ldots,x_N) = \sum_{i_1,\ldots,i_N \in \{0,1\}}
  c_{i_1,\ldots,i_N} x_1^{i_1} \cdots x_N^{i_N},\]
  with $c_{i_1,\ldots,i_N} \in \Rset^q$ for all $i_1,\ldots,i_N \in
  \{0,1\}$, and using the convention that if $i_k = 0$, then
  $x_k^{i_k} \equiv 1$.
\end{definition}

\cite{Belta-TAC06} showed  that a multi-affine function $h$ on a rectangle $\RNab$ is uniquely defined by its values at the vertices, and inside the rectangle the function is a convex combination of its values at the vertices:
\begin{equation}
  \label{eq:ma_convex}
  \begin{array}{l}
  h(x_1,\ldots,x_N) = \sum_{v \in \VNset} \prod_{i=1}^N \\
\left( \frac{x_i-a_i}{b_i-a_i} \right)^{\xi_i(v_i)} \left(
\frac{b_i-x_i}{b_i-a_i} \right)^{1-\xi_i(v_i)} \cdot h(v).
\end{array}
\end{equation}
where $\xi_i : \{a_i,b_i\} \longrightarrow \{0,1\}$ is an indicator function such that $\xi_i(a_i)=0$ and $\xi_i(b_i)=1$ for all $i=1,\ldots, N$.

%%%%%%%%%%%%%%%%%%%%%%%%%%%%%%%%%%%%%%%%%%%
\section{Problem formulation}\label{sec:prob}

Consider a continuous-time multi-affine control system of the form
\begin{equation}
  \label{eq:system}
  \dot{x}(t) = h(x(t)) + Bu(t), \hspace*{0.5cm} x(t) \in \RNset, u(t) \in U
\end{equation}
where $\RNset \subset \Rset^N$, $B \in \Rset^{N \times m}$, and the control input $u(t)$ is restricted to a polyhedral set $U \subset \Rset^m$.

Rectangular regions of interests in $\RNset$ are defined using a set of atomic propositions $\PREDSET = \{\pi_i \mid i=0, \ldots, l\}$. Each atomic proposition $\pi_i$ is satisfied in a set of  rectangular subsets of the state space of system~\eqref{eq:system}, which is denoted as:
\begin{equation}\label{eq:atomic_prop}
[\pi_i] = \cup_{j=1}^{d_i} R_N(a^{j,\pi_i},b^{j,\pi_i}) \subset \RNset, \hspace{10pt} d_i \in \Nset. 
\end{equation}

\begin{figure}
\centering
\includegraphics[width=0.5\columnwidth]{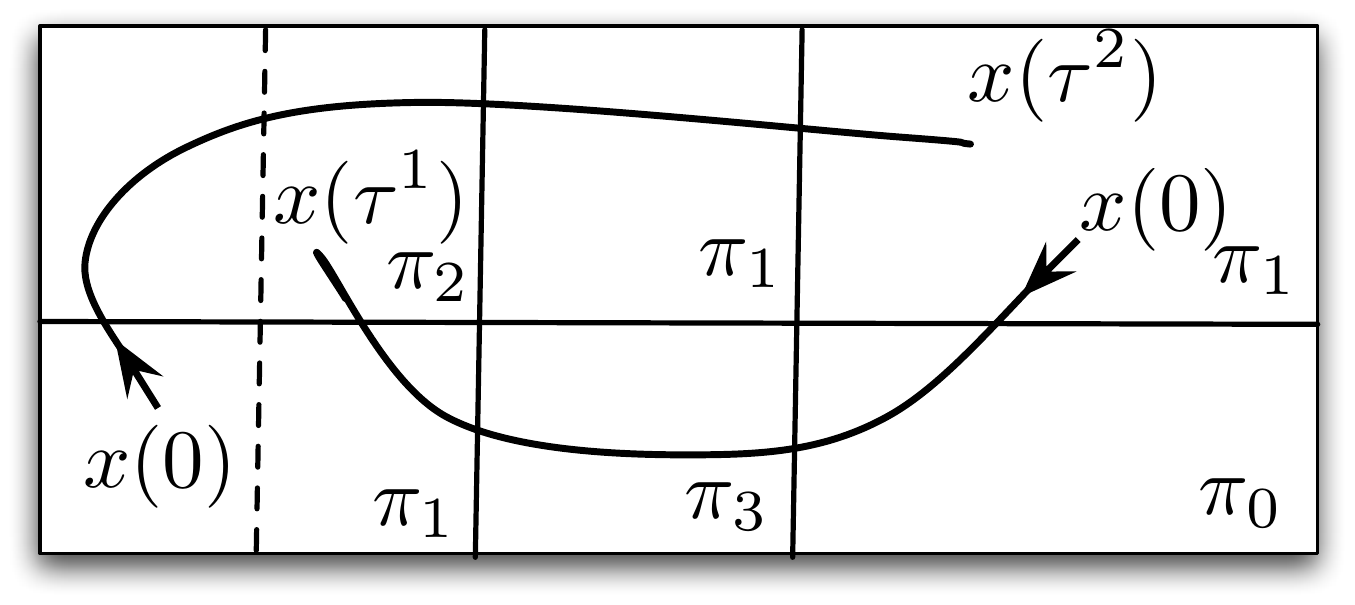}
\caption{Examples of continuous trajectories of system~\eqref{eq:system}. The atomic propositions are shown in the rectangles where they are satisfied.}
\label{fig:trajectory}
\end{figure}

The specifications are given as scLTL formulas over the set of predicates $\PREDSET$. A trajectory of system~\eqref{eq:system} satisfies the specification if the word produced by the trajectory satisfies the corresponding formula. 
Informally, while a trajectory of system~\eqref{eq:system} evolves, it produces the satisfying predicates and the sequence of predicates defines the word produced by a trajectory. Specifically, a trajectory produces predicate $\pi_i$ whenever it spends a finite amount of time in a rectangle where $\pi_i$ is satisfied. 
For example, trajectories $\{x(t)\}_{0\leq t \leq \tau^1}$ and $\{x(t)\}_{0\leq t \leq \tau^2}$ shown in Fig.~\ref{fig:trajectory} produce the words $\pi_1\pi_0\pi_3\pi_1\pi_2$ and $\pi_1\pi_2\pi_1\pi_1$, respectively. 
The word produced by a trajectory depends on how the rectangles are defined. The presented approach employs a refinement procedure based on adding hyperplanes, which induces smaller rectangles that inherit the predicate.
For example, if the dashed line in Fig.~\ref{fig:trajectory} is added, the trajectory $\{x(t)\}_{0\leq t \leq \tau^2}$ produces $\pi_1\pi_2\pi_2\pi_1\pi_1$.
As discussed by \cite{Kloetzer:2008}, when LTL without next operator is considered,   $\pi_1\pi_2\pi_1\pi_1$ and $\pi_1\pi_2\pi_2\pi_1\pi_1$ satisfy the same set of LTL formulas.
\begin{remark}
In this paper, we study finite time trajectories of system~\eqref{eq:system}. 
When infinite time trajectories are of interest, invariant controllers can be considered as in \cite{Habets2006}.
\end{remark}

\begin{problem}\label{prob:main}
Given a syntactically co-safe LTL formula $\Phi$ over a set of predicates $\PREDSET$ and a time bound \TIMEBOUND,  find a set of initial states $\XO \subset \RNset$ 
and a feedback control strategy such that all words produced by the closed-loop trajectories of system~\eqref{eq:system} originating in $\XO$ satisfy the formula in time less than \TIMEBOUND.
\end{problem}

Our proposed solution to Prob.\ref{prob:main} starts with a proposition-preserving rectangular partition\footnote{We use the term ``partition" loosely in this paper. 
The rectangle boundaries are irrelevant, since due to the synthesized controllers the trajectories never slide along the boundaries.} of $\RNset$, i.e., each element of the partition is a rectangle $\RNab \subseteq R_N(a^{j,\pi_i},b^{j,\pi_i})$ for some $j=1,\ldots,d_i$, $i=0,\ldots,l$ from Eqn.~\eqref{eq:atomic_prop}.
For each rectangle in the partition, and for each subset of its set of facets, we derive state-feedback controllers driving all the initial states in the rectangle through the set of facets in finite time by using the sufficient conditions derived in \cite{Habets2006}. We compute upper bounds for these times and choose the feedback controllers that minimize the upper bounds for each rectangle and each set of exit facets. We then construct a weighted transition system, in which the states label the rectangles from the partition, the inputs label the controllers, and the weights capture the time bounds. We find an optimal run of this transition system that satisfies the formula by solving an optimal reachability 
problem on its product with an FSA that accepts the language satisfying the formula.
The rectangles corresponding to the initial states with costs less than $\TIMEBOUND$ compose the set $\XO$. In order to increase this set, we use an iterative refinement of  the partition based on a random optimization algorithm.

\section{Facet Reachability Problems}
\label{sec:control}

In this section, we focus on the derivation of the facet reachability 
controllers and their corresponding time bounds. We first summarize the 
sufficient conditions for facet reachability from \cite{Habets2006}:

\begin{theorem}\label{thm:facet_reachability}
Let $\RNab$ be a rectangle and $\Fset \subset \FNab$ be a non-empty subset of its facets. There exists a multi-affine feedback controller $k : \RNab \longrightarrow U$ such that all the trajectories of the closed-loop system~\eqref{eq:system} originating in $\RNab$ leave it through a facet from the set $\Fset$ in finite time if the following conditions are satisfied:
\begin{align}
  & n_{F}^\top(h(v) + Bk(v)) \leq 0 ,  \forall  F \in {\cal F}_v \setminus \Fset \hspace{10pt} \forall v\in \VNset,  \label{eq:neg_speed}\\
  & 0 \not \in Conv( \{  h(v) + Bk(v) \mid v \in \VNset \}) \label{eq:nozero}
\end{align}
where $Conv$ denotes the convex hull.
\end{theorem}

In particular, when the cardinality of $\Fset$ is 1, i.e. $\Fset = \{F\}$, 
then Eqns. (\ref{eq:neg_speed}) and (\ref{eq:nozero}) imply that the speed towards the exit facet $F$ has to be positive everywhere in $\RNab$, i.e.

\begin{equation} \label{eq:pos_speed}
	0 < n_F^\top(h(v) + Bk(v)), \forall v\in \VNset.
\end{equation}
As a consequence, for this particular case, the sufficient conditions (\ref{eq:neg_speed}) and (\ref{eq:nozero}) can be replaced with (\ref{eq:neg_speed}) and 
(\ref{eq:pos_speed}).

The linear inequalities given in \eqref{eq:neg_speed} and \eqref{eq:pos_speed} (or  \eqref{eq:neg_speed} and  \eqref{eq:nozero}) 
define a set of admissible controls $U_v$ for each vertex $v \in \VNset$. By choosing a control for each vertex $v$ from the corresponding set $U_v$, we can construct a multi-affine state feedback controller $k$ that solves the corresponding control problem by using Eqn. (\ref{eq:ma_convex}). We first provide a time upper bound for the case when there is only one exit facet (Prop. \ref{prop:time_bound}), and then use this result to provide an upper bound for the general case (Cor. \ref{prop:multi_facet_time_bound}). 

\begin{proposition}\label{prop:time_bound}
Assume that $k: \RNab\longrightarrow U$ is an admissible multi-affine feedback controller that solves the control-to-facet problem for a facet $F \in \FNab$ with outer normal $e_i$ of a rectangle $\RNab$. Then all the trajectories of the closed loop system starting in rectangle $\RNab$ leave the rectangle through facet $F$ in time less than $T^F$, where
\begin{equation}\label{eq:TF}
		T^F = \ln( \frac{s_F}{\overline{s_F}})\frac{b_i - a_i}{  s_F   - \overline{s_F}},  \hspace{10pt}
  \end{equation}
with
\begin{align}
  	s_{F} = \minSAP \nonumber, \\
  	\overline{s_F} = \minSBP, \nonumber
  \end{align}
where $\overline {F}$ denotes the facet opposite to $F$, i.e. with normal $-e_i$.
\end{proposition}

{\bf Proof:} \newcommand{\lambdaP}{\ensuremath{\frac{b_i - x_i}{b_i - a_i}}} 
Let $x \in \RNab$ and $x^p, \overline{x^p}$ be the projections of $x$ on $F$ and $\overline{F}$, respectively. Then, we have $x = \lambdaP \overline {x^p} + (1- \lambdaP) x^p$.  
For every $x \in \RNab$, $h(x)$ is a convex combination of 
  $\{h(v) \mid v \in \VNset \}$. Furthermore, if $x$ belongs to
  a facet of $\RNab$, then $h(x)$ is a convex combination of the
  values of $h$ at the vertices of that facet. 
  Therefore, we have
  \begin{align}
  	s_{F}  \leq (h(x^p) + Bk(x^p))_i, \\
  	\overline{s_F}  \leq (h(\overline{x^p}) + Bk(\overline{x^p}))_i. 
  \end{align}
  Since $k(x)$ is a solution of the control-to-facet problem for facet $F$, the speed towards $F$ is positive everywhere in $\RNab$, hence
  \begin{equation}\label{eq:sx}
  	0 < s(x) := \lambdaP \overline{s_F} + (1-\lambdaP) s_F \leq (h(x) + Bk(x))_i.
  \end{equation}

  For any $x \in \RNab$, the speed in the $i^{th}$ direction is lower bounded by $s(x)$ (Eqn. ~\eqref{eq:sx}), which depends linearly on $x_i$.  Since system~\eqref{eq:slow} 
 defined below is always slower than the original one, its time upper bound to reach facet $F$ gives a valid upper bound for the original system.

	\begin{equation}\label{eq:slow}
		\dot{x_i}(t) = \lambdaP \overline{ s_F} + (1-\lambdaP) s_F, x_i \in [a_i, b_i]
	\end{equation}
	The explicit solution of Eqn.~\eqref{eq:slow} is given in Eqn.~\eqref{eq:solnslow}, where $x_i^0$ denotes the $i^{th}$ component of the initial condition.
	\begin{equation} \label{eq:solnslow}
		x_i(t) = \exp(  \frac{{s_F}_i - \overline{s_F}}{b_i - a_i}  t )(x_i^0 +  \frac{ b_i\overline{s_F }  - a_is_F} {s_F - \overline{s_F}  } ) -   \frac{ b_i\overline{s_F }  - a_is_F} {s_F - \overline{s_F}  } 
	\end{equation}
  	Solving~\eqref{eq:solnslow} for time $T^F_{x_i^0}$ at $x(T^F_{x_i^0}) = b_i$ gives the time upper bound from Eqn.~\eqref{eq:timeb}. Any trajectory starting from an initial point $x$ in $\RNab$ with $x_i = x_i^0$ reaches the facet $F^{e_i}$ in time less than $T^F_{x_i^0}$. 
	\begin{equation}\label{eq:timeb}
		T^F_{x_i^0} = \ln( \frac{ b_i + \frac{ b_i\overline{s_F}  - a_i s_F} {s_F - \overline{s_F}  } } { x_i^0 +  \frac{ b_i\overline{s_F}  - a_is_F} {s_F - \overline{s_F}  } } )\frac{b_i - a_i}{  s_F   - \overline{s_F}}
	\end{equation}
	As $T^F_{x_i^0}$ attains its maximum when $x_i^0 = a_i$,
	\begin{equation}
		T^F = \ln( \frac{ b_i + \frac{ b_i\overline{s_F }  - a_i s_F} {s_F - \overline{s_F}  } } { a_i +  \frac{ b_i\overline{s_F}  - a_i s_F} {s_F - \overline{s_F}  } } )\frac{b_i - a_i}{  s_F   - \overline{s_F}} = \ln( \frac{s_F}{\overline{s_F}})\frac{b_i - a_i}{  s_F   - \overline{s_F}}
	\end{equation}
	gives the upper bound for all $x \in \RNab$.
\qed
  
Prop.~\ref{prop:time_bound} uses the fact that if $k: \RNab \longrightarrow U$ is a solution to the considered control-to-facet problem, then the speed  $n_F^\top(h(x) + Bk(x))$ towards the exit facet is positive for all $x\in \RNab$. By defining a slower system using minimum speeds on $F$ and $\overline F$ towards the exit facet, a time bound for the original system is found. A more conservative time bound ${T'^F}$ can be computed using only the minimum speed towards $F$, i.e. ${T'^F} = \frac{b_i - a_i}{\min(s_F, \overline{s_F})}$. While it is more efficient to compute ${T'^F}$, $T^F$ gives a tighter bound ($T^F \leq {T'^F}$). Indeed, the computation of ${T^F}$ considers the change on the lower bound of speed with respect to $x_i$. Moreover, while $s_F$ gets closer to $\overline{s_F}$, $T'^F$ approaches $T^F$:
\begin{equation}
	\lim_{s_F \rightarrow \overline{s_F}} \ln( \frac{s_F}{\overline{s_F}})\frac{b_i - a_i}{  s_F   - \overline{s_F}} = \frac{ b_i - a_i}{\overline{s_F}}
\end{equation} 

\begin{remark} The time bound $T^F$ from Eqn.~\eqref{eq:TF} is attainable in some cases. Let $v_{s_F} =  \arg\min_{v \in \mathcal{V}(F)}{((h(v) + Bk(v))_i)}$ and $\overline{v_{s_F}}  = \arg\min_{v \in \mathcal{V}(\overline{F})}{((h(v) + Bk(v))_i)}$. If

\begin{equation}\label{eq:TF_ACHIEVE}
\begin{array}{l}
(v_{s_F})_j = (\overline{v_{{s_F}}})_j, \\
(h(v_{s_F}) + Bk(v_{s_F}))_j = 0, \\
(h(\overline{v_{s_F}}) + Bk(\overline{v_{s_F}}))_j=0, \hspace{10pt} j=1,\ldots,N, j \neq i,
\end{array}
\end{equation}
then the trajectory originating at  $\overline{v_{s_F}} \in \overline{F}$ reaches $v_{s_F} \in F$ at time $T^F$.
\end{remark}

For each vertex $v \in \VNset$, we can minimize the time bound 
given in Prop.~\ref{prop:time_bound} if we choose a control $u_v \in U_v$ that maximizes $n_F^\top(h(v) + Bu_v)$. Computationally, this involves 
solving a linear program at each vertex of a rectangle. Formally, at each vertex $v$, the optimization problem can be written as: 
\begin{align}
	& \ \ \ \ \ \ \ \ \ \ \ \ \max_{u_v}\ \ \  n_F^\top(h(v) + Bu_v) & \nonumber \label{eq:optim}\\
	&    n_{F'}^\top(h(v) + Bu_v) \leq -\epsilon ,  \forall  F' \in {\cal F}_v \setminus F \nonumber \\
	& u_v \in U
\end{align}

where $0 < \epsilon$, which is a robustness parameter guaranteeing  that a trajectory never reaches a facet other than $F$ while moving towards $F$. Decreasing $\epsilon$ relaxes the problem~\eqref{eq:optim} by increasing the size of the feasible region, which results in higher speeds and tighter time bounds.
Note that when $0 < \epsilon$ the equalities given in Eqn.~\eqref{eq:TF_ACHIEVE}  can not hold, 
since for a vertex $v$ the speed towards a facet $F' \in {\cal F}_v \setminus F$ is upper bounded by $-\epsilon$. 
Therefore the robustness parameter $\epsilon$ also affects the distance between the time bound from Eqn.~\eqref{eq:TF} and the actual maximal amount of time required to reach $F$. 
\begin{figure}
\centering
\subfloat[]{\label{fig:ex_rec}\includegraphics[width=0.5\columnwidth]{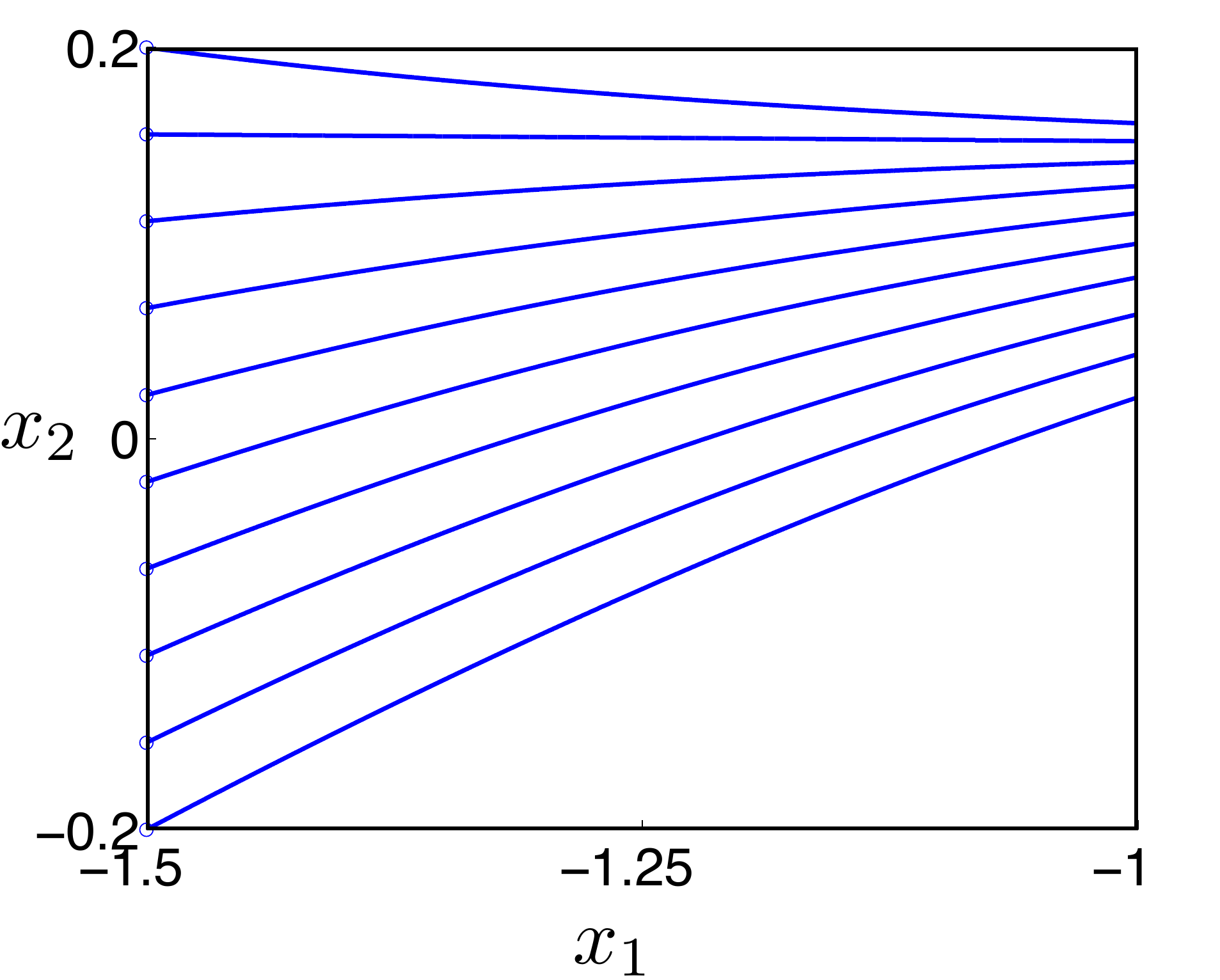}}
\subfloat[]{\label{fig:ex_recTime}\includegraphics[width=0.5\columnwidth]{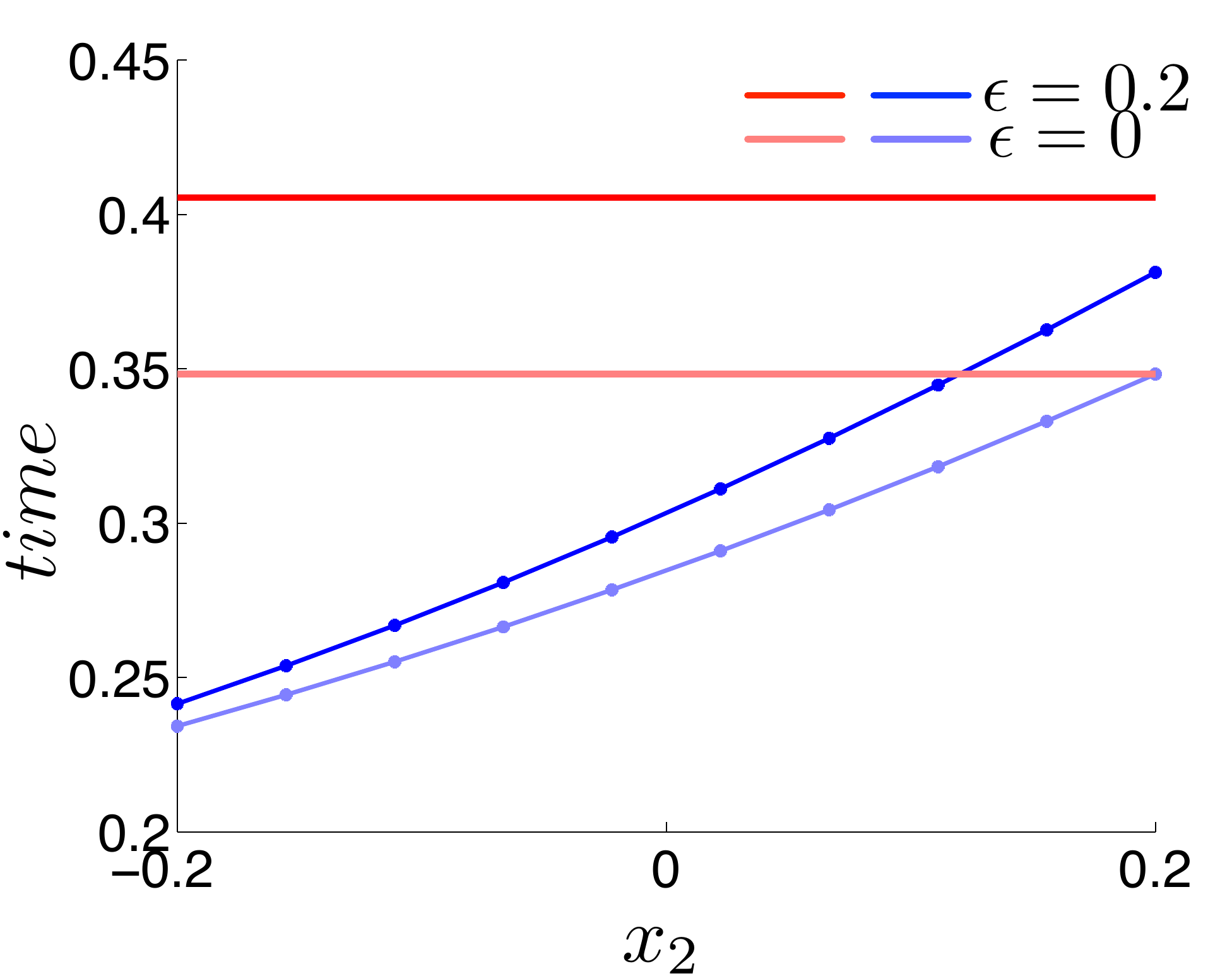}}
\caption{(a) Rectangle $[-1.5, -1] \times [-0.2, 0.2]$ and sample trajectories originating in facet $F^{-e_1}$. 
(b) Simulation times according to the initial condition in $x_2$ and time bounds $T^{F^{e_1}}$ (red lines) computed using
Eqn. (\ref{eq:TF}) for controllers synthesized from \eqref{eq:optim} for $\epsilon=0$ and $\epsilon=0.2$.}
\label{fig:ex_r}
\end{figure} 

The tightness of the time bound from Eqn.~\eqref{eq:TF} and the effects of the robustness parameter $\epsilon$ are illustrated through an example in Fig.~\ref{fig:ex_r}, where the control problem for exit facet $F^{e_1}$ of rectangle $[-1.5, -1] \times [-0.2, 0.2]$ is considered for the control system from Eqn. (\ref{eq:case_system}). Some trajectories of the closed loop system obtained by using the feedback controller that minimizes $T^{F^{e_1}}$ when $\epsilon=0.2$ are shown in Fig.~\ref{fig:ex_rec}. 
The corresponding times for reaching $F^{e_1}$ for $\epsilon=0$ and $\epsilon=0.2$ are shown in Fig.~\ref{fig:ex_recTime}. 
Note that when $\epsilon=0$, the trajectory starting from $(-1.5, 0.2)$ reaches facet $F^{e_1}$ exactly at time $T^{F^{e_1}}$.

\begin{corollary}\label{prop:multi_facet_time_bound}
Given a rectangle $\RNab$ and  an admissible multi-affine feedback $k: \RNab
\longrightarrow U$ that solves the control problem from Thm. \ref{thm:facet_reachability} with set of exit facets $\Fset$, all trajectories of the closed loop system originating in rectangle $\RNab$ leave it through a facet $F \in \Fset$ in time less than 
\[
T^{\Fset} = \min_{F \in \Fset} T^F,
\]
 where each $T^F$ is computed as in Prop.~\ref{prop:time_bound} if $0 < n_F^\top (h(v) + Bk(v)) $ for all $v \in \VNset$. Otherwise $T^F$ is set to $\infty$.
\end{corollary}

{\bf Proof:} Let $F \in \Fset$ with $0 < n^F (h(v) + Bk(v)) $ for all $v \in \VNset$. Then by Prop.~\ref{prop:time_bound} every trajectory originating in $\RNab$ reaches $F$ within time $T^F$~\eqref{eq:TF} unless it leaves $\RNab$ before reaching $F$. Hence, $\min_{F \in {\cal E}} T^F$ gives a valid bound to the control-to-set-of-facets problem for $\Fset$.
\qed

For a facet reachability problem with $\Fset$ as the set of exit facets,  $T^F$ is computed for each $F \in \Fset$ through choosing controls that minimize $T^F$~\eqref{eq:TF} and satisfy the linear inequalities defined in Thm.~\ref{thm:facet_reachability}. Computationally, this translates to solving the following linear program for each $v \in \VNset$ and for each $F \in \Fset$: 
\begin{align}
	& \ \ \ \ \ \ \ \ \ \ \ \ \max_{u_v}\ \ \  n_F^\top(h(v) + Bu_v) & \nonumber \label{eq:optimM}\\
	&    n_{F'}^\top(h(v) + Bu_v) \leq -\epsilon ,  \forall  F' \in {\cal F}_v \setminus \Fset \nonumber \\
	& u_v \in U
\end{align}
where $\epsilon$ is defined as in optimization problem \eqref{eq:optim}.

As already stated, $T^F$ for $F \in \Fset$ is calculated as in~\eqref{eq:TF} if the speeds at all vertices are positive towards $F$. In this case, the condition from~\eqref{eq:nozero} is trivially satisfied. Then a multi-affine feedback $k$ is constructed by using the controls where $\min_{F \in \Fset} T^F$ attains its minimum. 

 %%%%%%%%%%%%%%%%%%%%%%%%%%%%%%%%%%%%%%%%%%%
\section{Control Strategy}\label{sec:controlStrategy}

In this section, we provide a solution to Prob.~\ref{prob:main} for a proposition-preserving partition of $\RNset$. We use the results from Sec.~\ref{sec:control} to construct a weighted transition system from the partition and find an optimal control strategy for the weighted transition system. The control strategy enforces the satisfaction of the specification and maps directly to a strategy for system~\eqref{eq:system}.

A proposition-preserving partition of $\RNset$ and solutions of facet reachability problems for the rectangles in the partition set define a weighted transition system $\TS=(Q, \Sigma, \delta, O, o, w)$.
Each state $q\in Q$ of $\TS$ corresponds to a rectangle $R_N(a^q, b^q)$ in the partition set. 
An input $\sigma \in \Sigma$ of $\TS$ indicates a non-empty subset of the facets of a rectangle and a transition $\delta(q, \sigma)$ is introduced if the corresponding control problem has a solution. 
Specifically, we consider a facet reachability problem for each state $q \in Q$ and each non-empty subset of ${\cal F}(a^q,b^q)$, and find the multi-affine feedback control which minimizes the corresponding time bound as explained in Sec.~\ref{sec:control}.
The successors of  $\delta(q, \sigma)$ are the states $q'$ such that $R_N(a^q, b^q)$ and $R_N(a^{q'}, b^{q'})$ have a common facet in $\sigma$. 
The transition weights are assigned according to the time bounds computed as described in Prop.~\ref{prop:time_bound} and Cor.~\ref{prop:multi_facet_time_bound}. $O$ equals to the set of predicates $\PREDSET$ and $o(q) = \pi_i$ if $R_N(a^q,b^q) \subseteq [\pi_i]$.

All words that satisfy the specification formula $\Phi$ are accepted by a FSA $\AUTOMATON = (S, \PREDSET, \delta_\AUTOMATON, S_0, F)$ \footnote{In the general case, as described in Sec. \ref{sec:pre}, the input alphabet of this automaton is $2^\PREDSET$. However, since the words generated by system (\ref{eq:system})
are over $\PREDSET$, it is sufficient to consider $\PREDSET$ as the input alphabet for the automaton.}. We construct a product automaton  $\AUTOMATON^P = (S_P, \Sigma, \delta_P, S_{P0}, F_P)$ from $\TS$ and $\AUTOMATON$ as described in Def.~\ref{def:product_automaton}.

\newcommand{\ControlSet}{{SC}}
\newcommand{\StateCost}{J}

A control strategy $(S_\Omega, \Omega)$ for $\AUTOMATON^P$ is defined as a set of initial states $S_\Omega$ and a state feedback control function $\Omega : S_P \longrightarrow \Sigma$ implying that $\Omega(s)$ will be the input at state $s$. The state feedback function $\Omega$ characterizes the set of initial states $S_{\Omega} \subset S_{P0}$ such that every run $s_0s_1\ldots s_n$ of $\AUTOMATON^P$ starting from a state $s_0$ in $S_{\Omega}$ is an accepting run over the word $\Omega(s_0)\ldots \Omega(s_{n-1})$.
Since $\AUTOMATON^P$ is non-deterministic, there can be multiple runs starting from a state $s_0 \in S_\Omega$ under the feedback control $\Omega$. In literature (\cite{Kloetzer:2008NonDet}, \cite{Wolfgang2002}), non-determinism is resolved through a reachability game played between a protagonist and an adversary, and  $S_\Omega$ is defined as the set of initial states such that the protagonist always wins the game by applying $\Omega$. 
Next, we introduce an algorithm based on fixed-point computation to find a maximal $S_\Omega$ and corresponding feedback control $\Omega$ through optimizing a cost for each $s \in S_P$. \cite{Asarin:1999} used a similar algorithm to solve optimal reachability problems on timed game automata. 

\begin{remark}
Generally, the reachability games are considered over an infinite horizon such as Buchi games, where winning a game for the protagonist means identifying and reaching an invariant set of ``good'' states.  
As we consider FSAs, the acceptance condition coincides with finite time reachability. 
Hence, a simple reachability algorithm is sufficient in our case.
\end{remark}

Let $\StateCost_{\Omega}:S_P \rightarrow \Rset_+$ be a cost function with respect to a set of final states $F_P$ and feedback control $\Omega$ such that any run of $\AUTOMATON^P$ starting from $s$ reaches a state $f \in F_P$ under the feedback control $\Omega$ with a cost upper bounded by  $\StateCost_{\Omega}(s)$. Note that if there exists a run starting from $s$ that can not reach $F_P$, the cost is infinity, $\StateCost_{\Omega}(s) = \infty$.

The solution of the fixed-point problem given in Eqn. (\ref{eq:fixed_point}) gives the optimal cost for each $s\in S$.
	\begin{equation}\label{eq:fixed_point}
		\StateCost(s) = \min(\StateCost(s) ,   \min_{\sigma \in \Sigma} \max_{s' \in \delta_P(s,\sigma)} \StateCost(s') + w_P(\delta_P(s,\sigma)) )
	\end{equation}

\begin{algorithm}\caption{\small{Compute $\StateCost$ and $\Omega$ for  $\AUTOMATON^P = (S_P, \Sigma, \delta_P, S_{P0}, F_P)$}}
\label{algo:fixed_point}
\small{
\begin{algorithmic}[1]
	\State $\StateCost(s) = \infty, \forall s\in S_P$
	\State $\StateCost(f) = 0, \forall f \in F_P$
	\State $\ControlSet = \{ s | \exists \sigma \in \Sigma$ and $f \in F_P$ such that $f \in \delta_P(s,\sigma)\}$
	\While{ $ \ControlSet \neq \emptyset$}
		\State $\ControlSet = \ControlSet \setminus \{ s\}, for\ some\ s \in \ControlSet$
			\If{$\min_{\sigma \in \Sigma} \max_{s' \in \delta_P(s,\sigma)} \StateCost(s') + w_P(\delta_P(s,\sigma))   < \StateCost(s) $} \label{alg:update_rule}
				\State $\Omega(s) = \arg \min_{\sigma \in \Sigma} \max_{s' \in \delta_P(s,\sigma)} \StateCost(s') + w_P(\delta_P(s,\sigma))$			
				\State $\StateCost(s) =  \max_{s' \in \delta_P(s, \Omega(s))} \StateCost(s') +  w_P(\delta_P(s,\Omega(s)))$
				\State $\ControlSet = \ControlSet \cup \{ s' | \exists \sigma \in \Sigma, s \in \delta_P(s',\sigma)\}$
			\EndIf
	\EndWhile
\end{algorithmic}
}
\end{algorithm}

Alg.~\ref{algo:fixed_point} implements the solution for the fixed-point problem in Eqn.~\eqref{eq:fixed_point} for the states of $\AUTOMATON^P$
and finds the optimal feedback control $\Omega$. 
A finite state cost, $\StateCost(s)< \infty$, and a feedback control $\Omega$ resulted from Alg.~\ref{algo:fixed_point} means that every run starting from $s$ reaches a state $f$ in $F_P$ under the feedback control $\Omega$ with a cost at most $\StateCost(s)$. Therefore, $S_{\Omega} = \{ s \mid \StateCost(s) < \infty ,  s \in S_0\}$ is the maximal set of initial states of $\AUTOMATON^P$ such that under the feedback control $\Omega$ all runs starting from $S_{\Omega}$ are accepting. 
Consequently, 
\begin{equation}\label{eq:s_omegaT}
S_{\Omega}^T = \{ s \mid \StateCost(s) < T ,  s \in S_0\}
\end{equation} 
is the maximal set of initial states such that under the feedback control $\Omega$ cost of a run starting from $S_{\Omega}^T$ is upper bounded by $T$.

If only control-to-facet problems are considered while constructing the transition system $\TS$, $\TS$ and the product automaton $\AUTOMATON^P$ become deterministic. Hence, in this case it is sufficient to use a shortest path algorithm to find optimum costs and feedback control $\Omega$ instead of Alg.~\ref{algo:fixed_point}.

If a multi-affine feedback $k$ solves facet reachability problem for the set of exit facets $\Fset \subset \FNab$ of rectangle $\RNab$, then $k$ is a solution of the facet reachability problem for every superset $\Fset'$ of $\Fset$ with the same time bound $T^\Fset$ by Cor.~\ref{prop:multi_facet_time_bound}. While constructing $\delta$ of $\TS$, a solution is searched for every subset of $\FNab$, hence
\begin{equation}
	w_P(\delta_P(s, \Fset')) \leq w_P(\delta_P(s,\Fset)), if \Fset \subseteq \Fset'.
\end{equation} 
In line~ \ref{alg:update_rule} of Alg.~\ref{algo:fixed_point}, cost of a state is updated according to the state with maximum cost among a transitions successor states, hence Alg.~\ref{algo:fixed_point} tends to choose the $\Fset$ with minimum cardinality among the sets $\Fset' \subset \FNab$ with the same transition cost.

\textbf{\emph{Control Strategy for $\TS$: }} (\cite{Kloetzer:2008NonDet})
We construct a control strategy $(Q_0,\AUTOMATON^C)$ for $\TS$ using the control strategy $(S^\TIMEBOUND_\Omega, \Omega)$ for $\AUTOMATON^P$ resulted from Alg.~\ref{algo:fixed_point} and Eqn.\eqref{eq:s_omegaT}.
The set of initial states $Q_0$ is the projection of $S^\TIMEBOUND_\Omega$ to the states of $\TS$.
Since the feedback control $\Omega$ for $\AUTOMATON^P$ becomes non-stationary when projected to the states of $\TS$, we construct a feedback control for $\TS$ in the form of a feedback control automaton $\AUTOMATON^C = (S_C, Q, \delta_C, S_{C0}, F_C, \Omega_C, \Sigma)$. 
The feedback control automaton $\AUTOMATON^C$ reads the current state of $\TS$ and outputs the input to be applied to that state. 
The set of states $S_C$, the set of initial states $S_{C0}$ and the set of final states $F_C$ of $\AUTOMATON^C$ are inherited from $\AUTOMATON$, the set of inputs $Q$ is the states of $\TS$. 
The memory update function $\delta_C : S_C \times Q \longrightarrow S_C$ is defined as  $\delta_C(s,q) = \delta_\AUTOMATON(q,o(q))$ if $\delta_\AUTOMATON(q,o(q))$ is defined. The output alphabet $\Sigma$ is the input alphabet of $\TS$. $\Omega_C : S_C \times Q \longrightarrow \Sigma$ is the output function, $\Omega_C(s,q) = \Omega((q,s))$ if $J((q,s)) < T$ and $\Omega_C(s,q)$ is undefined otherwise.

If we set the set of observations of $\TS$ to $Q$ and define the observation map $o$ as an identity map, then the product of $\TS$ and $\AUTOMATON^C$ will have same states and transitions as $\AUTOMATON^P$. 
Hence, the words produced by trajectories of $\TS$ starting from $Q_0$ in closed loop with $\AUTOMATON^C$ satisfy $\Phi$.

Control strategy $(Q_0,\AUTOMATON^C)$ for $\TS$ is used as a control strategy for system~\eqref{eq:system} by mapping the output of $\AUTOMATON^C$ to the corresponding multi-affine feedback controller. 
This strategy guarantees that every trajectory of system~\eqref{eq:system} originating in $\XO$ given in Eqn.~\eqref{eq:X0} satisfies $\Phi$ in time less than \TIMEBOUND.
\begin{equation}\label{eq:X0}
	\XO = \bigcup_{q \in Q_0} R_N(a^q, b^q)
\end{equation} 
For every $x_0 \in \XO$, there exists an initial state $q \in Q_0$ and $s \in S_{C0}$ such that $x_0 \in R_N(a^q, b^q)$ and $(q,s) \in S_\Omega^T$ from Eqn.~\eqref{eq:s_omegaT}.
Let $k_{s,\Omega_C(s,q)}$ be the multi-affine feedback which solves control-to-facet (or control-to-set-of-facets) problem on $R_N(a^q, b^q)$ for $\Omega_C(s,q)$ as the set of exit facets. Starting from $x_0$ multi-affine feedback $k_{s,\Omega_C(s,q)}$ is applied to system~\eqref{eq:system} until the trajectory reaches a facet $F \in \Omega_C(s,q)$ with a positive speed towards $F$. By construction of $\AUTOMATON^C$, it is guaranteed that the trajectory reaches a facet $F \in \Omega_C(s,q)$ in time less than $w(\delta(q, \Omega_C(s,q)))$.
Then the applied multi-affine feedback switches to $k_{s',\Omega_C(s',q')}$ where $F = R_N(a^q, b^q) \cap R_N(a^{q'}, b^{q'})$ and $s' = \delta_C(s,q)$. This process continues until a final state $f \in F_C$ of $\AUTOMATON^C$ is reached.

\begin{theorem}\label{thm:main}
The trajectories of system~\eqref{eq:system} originating in $\XO$~\eqref{eq:X0} with control strategy $(Q_0,\AUTOMATON^C)$ satisfies $\Phi$ in time less than \TIMEBOUND.  
\end{theorem}	
{\bf Proof:}
By Def.~\ref{def:product_automaton}, every word produced by an accepting run of $\AUTOMATON^P$ satisfies $\Phi$. Hence, by construction of $(Q_0, \AUTOMATON^{C})$ and $\XO$ the words produced by closed loop trajectories of system~\eqref{eq:system} originating in $\XO$ satisfy $\Phi$.
Consider a finite trajectory $\{x(t)\}_{0 \leq t \leq \tau}$ of system~\eqref{eq:system} with $x(0) \in \XO$ evolving under the control strategy  $(Q_0,\AUTOMATON^C)$.
Let  $r_C = s_0s_1\ldots s_n$ be the corresponding run of $\AUTOMATON^C$,  $r_{\TS} = q_0q_1\ldots q_n$ be the corresponding trajectory of $\TS$ and $t_i$ be a time instant when control switch occurs, i.e. at time $t_i$, the trajectory hits a facet $F \in \Omega_C(s_{i-1},q_{i-1})$ with a positive speed towards $F$ while evolving under the multi-affine feedback $k_{s_{i-1}, \Omega_C(s_{i-1},q_{i-1})}$, for all $i=1,\ldots,n$ and $t_{n}=\tau$.
By Prop.~\ref{prop:time_bound} and Cor.~\ref{prop:multi_facet_time_bound}, for all $i=1,\ldots,n$:
\begin{equation}
t_i - t_{i-1} \leq w(\delta(q_{i-1}, \Omega_C(s_{i-1}, q_{i-1}))) = T^{\Omega_C(s_{i-1}, q_{i-1})}.
\end{equation}
By Alg.~\ref{algo:fixed_point}, $\tau \leq \StateCost((q_0,s_0))$ and by Eqn.~\eqref{eq:X0} $\tau \leq \TIMEBOUND$.
\qed

In Thm.~\ref{thm:main}, we showed that the proposed feedback control strategy solves Prob.~\ref{prob:main} for a proposition-preserving partition of $\RNset$.  Next we describe an iterative refinement procedure to increase the volume of $\XO$.

 %%%%%%%%%%%%%%%%%%%%%%%%%%%%%%%%%%%%%%%%%%%

\section{Refinement}\label{sec:optimization}
 \newcommand{\OPTVAR}{d_{i*}^{j}}
 \newcommand{\OPTVARS}{d_{i}^{j}}
 \newcommand{\OPTVARN}{d_{i+1}^{j}}
\newcommand{\NEWTH}{d^j_*}

An iterative refinement procedure is employed to enlarge the set $\XO$~\eqref{eq:X0}. 
As mentioned before, the rectangles defined by the set of predicates induce an initial proposition-preserving grid partition of $\RNset$. A grid partition is defined by a set of thresholds $\{d_i^j\}_{i \in \Nset}$ for each dimension $1\leq j \leq N$.

\begin{figure}
\centering
\includegraphics[width=0.8\columnwidth]{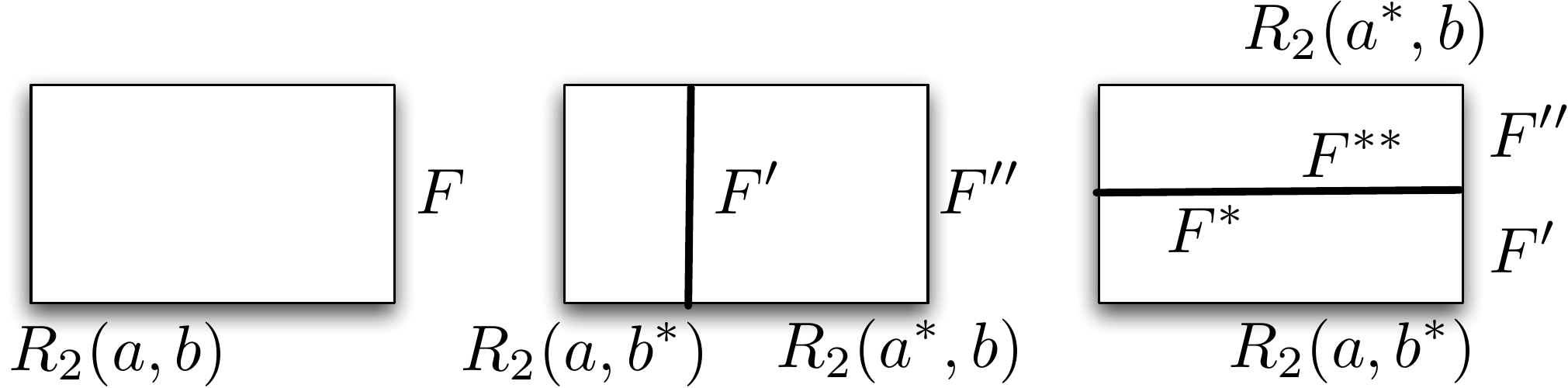}
\caption{Two partitioning schemes for $R_2(a.b) \subset \Rset^2$.}
\label{fig:partition}
\end{figure}

 Introducing a new threshold $\NEWTH$ in dimension $j$ can affect $\XO$ in different ways and it does not always enlarge the set $\XO$.  Consider a state $s \in S_{P0}$ with $\StateCost(s)$ as computed in Alg.~\ref{algo:fixed_point} and corresponding rectangle $\RNab$ with ${a}_j < \NEWTH < {b}_j$.  Assume a multi-affine feedback $k: \RNab
\longrightarrow U$ solves the control-to-facet problem for a facet $F \in \FNab$ with outer normal $e_i$ and assume the corresponding time bound is $T^F$ as given in Prop.~\ref{prop:time_bound}. When $\RNab$ is partitioned into two rectangles $R_N(a , b^*)$ and $R_N(a^* , b)$ through a hyperplane $x_j = d^j_*$,  we need to consider two cases: $j=i$ and $j\neq i$, which are illustrated in Fig.~\ref{fig:partition} on a rectangle in $\Rset^2$.

  \emph{(a)}\{{\it $i=j$}\} Since state feedback $k$ solves the control-to-facet problem on $\RNab$ for $F$, the speed towards the exit facet is positive for all $x \in \RNab$. Moreover, no trajectory leaves $\RNab$ through another facet. Hence, $k$ solves the control-to-facet problems on $R_N(a,b^*)$ and $R_N(a^*,b)$ for the facets with normal $e_i$. Let $T^{F'}$ and $T^{F''}$ be the corresponding time bounds. Then when $k$ is applied, any trajectory starting in $R_N(a,b^*)$ and $R_N(a^*,b)$ reaches $F$ within time $T^{F'} + T^{F''}$, which is upper bounded by $T^F$. The proof follows from the proof of the Prop.~\ref{prop:time_bound}, the minimal speed towards $F$ on the intersection of $\RNab$ and $x_i = \NEWTH$ is lower bounded by $\frac{b_i - \NEWTH}{b_i - a_i}\overline{s_{F}} + \frac{\NEWTH - a_i}{b_i - a_i}s_{F}$. As the actual minimal speed could be higher than $\frac{b_i - \NEWTH}{b_i - a_i}\overline{s_{F}} + \frac{\NEWTH - a_i}{b_i - a_i}s_{F}$ and other multi-affine feedbacks could solve the same problem on $R_N(a,b^*)$ and $R_N(a^*,b)$ with lower time bounds, when $i=j$, partitioning results in tighter time bounds.

  \emph{(b)}\{{\it $i\neq j$}\} The multi-affine feedback $k$ solves the control-to-set-of-facets problem on $R_N(a,b^*)$ for exit facets $\Fset^1 = \{F' , F^*\}$ where $n_{F'} =  e_i$ and $n_{F^*} = e_j$. Moreover, $k$ solves control-to-set-of-facets problem on  $R_N(a^*,b)$ for exit facets $\Fset ^2= \{F'' , F^{**}\}$ where $n_{F''} =  e_i$ and $n_{F^{**}} = -e_j$. Then the corresponding time bounds $T^{\Fset^1}$ and  $T^{\Fset^2}$ are upper bounded by $T^F$ by Cor.~\ref{prop:multi_facet_time_bound}. However, $T^{F'}$ or $T^{F''}$ could be higher than $T^F$, hence, the costs of the resulting automaton states could be higher than $\StateCost(s)$.

In \emph{(a)} and \emph{(b)}, the effects of partitioning are analyzed on a rectangular region for a simple case where the initial rectangle has a solution to the control-to-facet problem for facet $F$. It is concluded that when a rectangle  $\RNab$ of a state $s \in S_{P}$ with $\StateCost(s)$  is partitioned, the costs of the resulting states $s'$ and $s''$ can be higher or lower than $\StateCost(s)$. 
Hence, even for that simple case, partitioning can have negative and positive effects on the defined time bound for a single rectangle.
Moreover, there is no closed form relationship between the partitioning scheme $\{d_i^j\}_{i \in \Nset}$  and the volume of the set $\XO$.

In order to overcome these difficulties,
we use a Particle Swarm Optimization (PSO)(\cite{PSOTralea}) algorithm to find the new thresholds. The objective of the optimization is maximizing the volume of the set $\XO$~\eqref{eq:X0}. We run the PSO algorithm iteratively.
At each iteration, a new threshold $\OPTVAR$ is added between two consecutive ones $\OPTVARS, \OPTVARN$ depending on the distance between them and the value of the corresponding optimization variable. An optimization variable for $\OPTVAR$ is defined with range $[\OPTVARS, \OPTVARN - d]$ if the distance between two consecutive thresholds is twice as large as the minimum allowed edge size, $2d < \OPTVARN- \OPTVARS$. Part of the range $[\OPTVARS, \OPTVARS + d)$ is used to decide whether to add the threshold or not, i.e. a new threshold is added only if $\OPTVAR \in [\OPTVARS + d,\OPTVARN-d]$. 
The dimension of the optimization problem depends on the grid configuration $\{d_i^j\}_{i \in \Nset}$ of the iteration.
The iterative procedure terminates when either all the intervals are smaller than $2d$ or there is no change in the optimum objective value for the last two iterations. 

\begin{remark} Let $d^j = | \{d^j_i\}_{i \in \Nset}|$, then the cardinality of the resulting partition is $\Pi_{j=1}^N (d^j - 1)$.
Construction of the transition system $\TS$ (see Sec.~\ref{sec:controlStrategy}) from the partition $\{d^j_i\}_{i \in \Nset}$ requires to solve $(2^{2N} - 1)\Pi_{j=1}^N (d^j - 1)$ linear programs. For each partition, in addition to solving these linear programs, we take the product between $\TS$ and $\AUTOMATON$, and run Alg.~\ref{algo:fixed_point} to find the volume of the set $\XO$.
\end{remark}

 %%%%%%%%%%%%%%%%%%%%%%%%%%%%%%%%%%%%%%%%%%%
\section{Case Study}\label{sec:casestudy}

\begin{figure}
\centering
\subfloat[]{\label{fig:ex_fsa}\includegraphics[width=0.55\columnwidth]{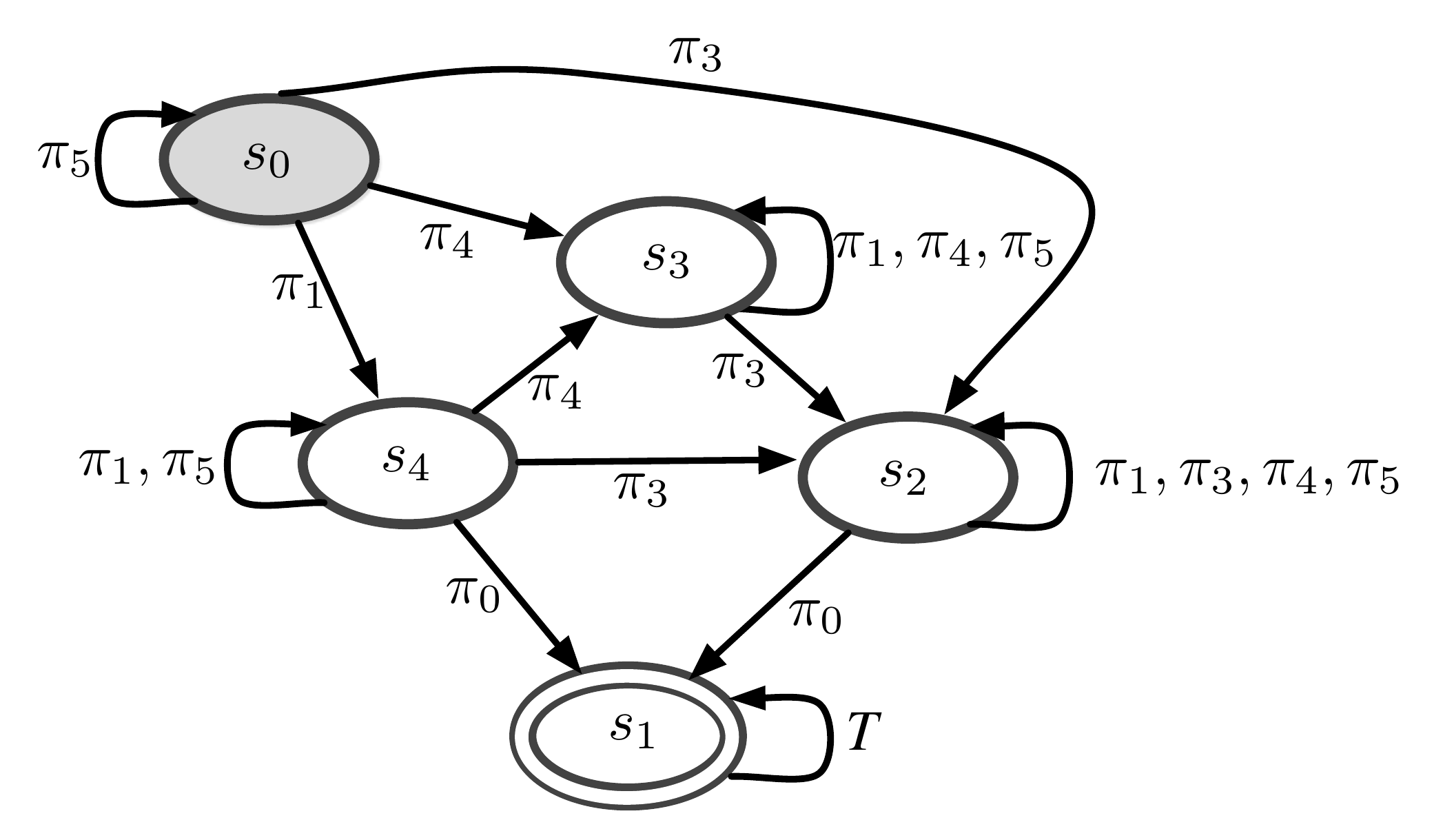}}
\subfloat[]{\label{fig:ex_env}\includegraphics[width=0.35\columnwidth]{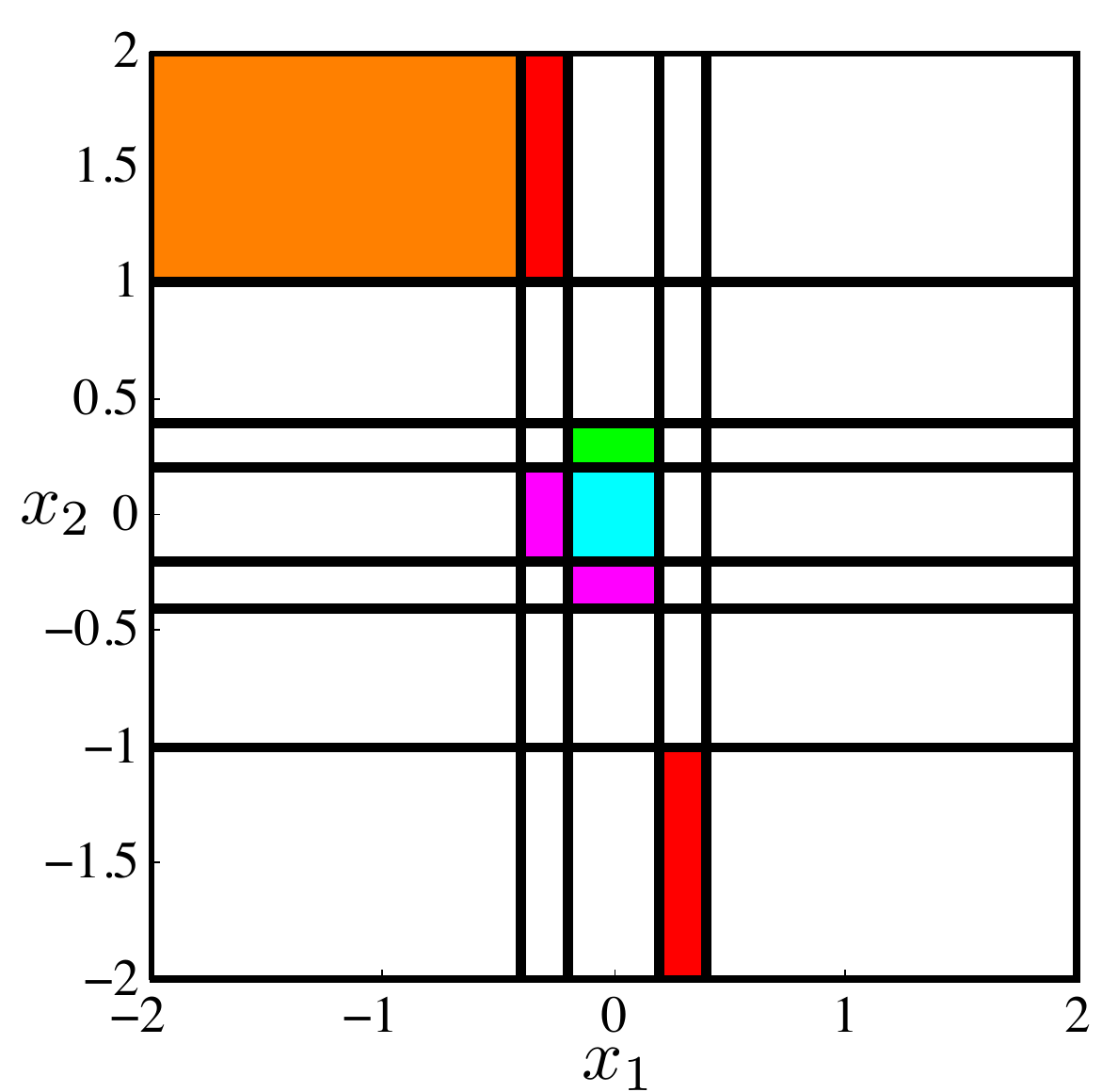}}
\caption{(a) FSA $\AUTOMATON$ that accepts the language satisfying $\Phi$~\eqref{eq:spec_case_study} ($T$ stands for Boolean constant true). The initial state of the automaton is filled with grey and the final state is marked with a double circle. (b) The initial partition induced by the predicate set $\PREDSET$. $\pi_0$, $\pi_1$, $\pi_2$, $\pi_3$ and $\pi_4$ are satisfied in cyan, magenta, red, green and orange colored rectangles, respectively and $\pi_5$ is satisfied in white rectangles.}
\end{figure}

%%%%%%%%%%%%%%%%%%%%%%%%%%%%%%%%%%%%%%%%%%%
Consider the following multi-affine system
\begin{equation}\label{eq:case_system}
\begin{array}{ccc}
	\dot x_1 & = & -x_1 + x_1x_2 + u \\
	\dot x_2 & = & -x_2 + x_1x_2 + u,	 
\end{array}
\end{equation}
where the state $x$ and the control input $u$ are constrained to sets 
$\RNset = [-2,2] \times [-2,2]$ and $U = [-1,1]$, respectively.
The specification is to visit one of the rectangles that satisfy $\pi_1$ or $\pi_3$, then a rectangle where $\pi_0$ is satisfied, while always avoiding the rectangles that satisfy $\pi_2$. Moreover, if a trajectory visits a rectangle where $\pi_4$ is satisfied, then it has to visit a rectangle that satisfies $\pi_3$ before visiting a rectangle that satisfies $\pi_0$. Predicates $\pi_i$, $i=0,\ldots,4$ are defined in Fig. \ref{fig:ex_env}. Formally, this specification translates to the following scLTL formula $\Phi$ over $\PREDSET = \{\pi_0,\pi_1,\pi_2,\pi_3, \pi_4, \pi_5\}$: 
\begin{equation}\label{eq:spec_case_study}
	\Phi = ( ( \neg \pi_4 \LTLUNTIL \pi_0 ) \vee ( \neg \pi_0 \LTLUNTIL \pi_3) ) \wedge ( \neg \pi_2 \LTLUNTIL \pi_0 ) \wedge ( \neg \pi_0 \LTLUNTIL ( \pi_1 \vee \pi_3 ) )
\end{equation}

A FSA $\AUTOMATON$ that accepts the language satisfying formula $\Phi$ is given in Fig.~\ref{fig:ex_fsa}. The regions of interests and the corresponding partition are given in Fig.~\ref{fig:ex_env}. The upper time bound to satisfy the specification is set to $\TIMEBOUND=2.5$, the minimum edge length is set to $d=0.2$ and the robustness parameter for optimization problems~\eqref{eq:optim} and~\eqref{eq:optimM} is set to $\epsilon=0.2$.

To illustrate the main results of the paper, we use two approaches to generate a control strategy. In the first experiment, only control-to-facet problems are considered, hence a deterministic transition system is used.
As discussed in the paper, the resulting product automaton is also deterministic and it is sufficient to use a shortest path algorithm instead of Alg.~\ref{algo:fixed_point}. 
In the second approach, both control-to-facet and control-to-set-of-facets problems are considered. Hence, the resulting transition system and product automaton are non-deterministic, and Alg.~\ref{algo:fixed_point} is applied.

\begin{figure}[h]
\centering

\subfloat[]{\label{fig:ex_det}\includegraphics[width=0.45\columnwidth]{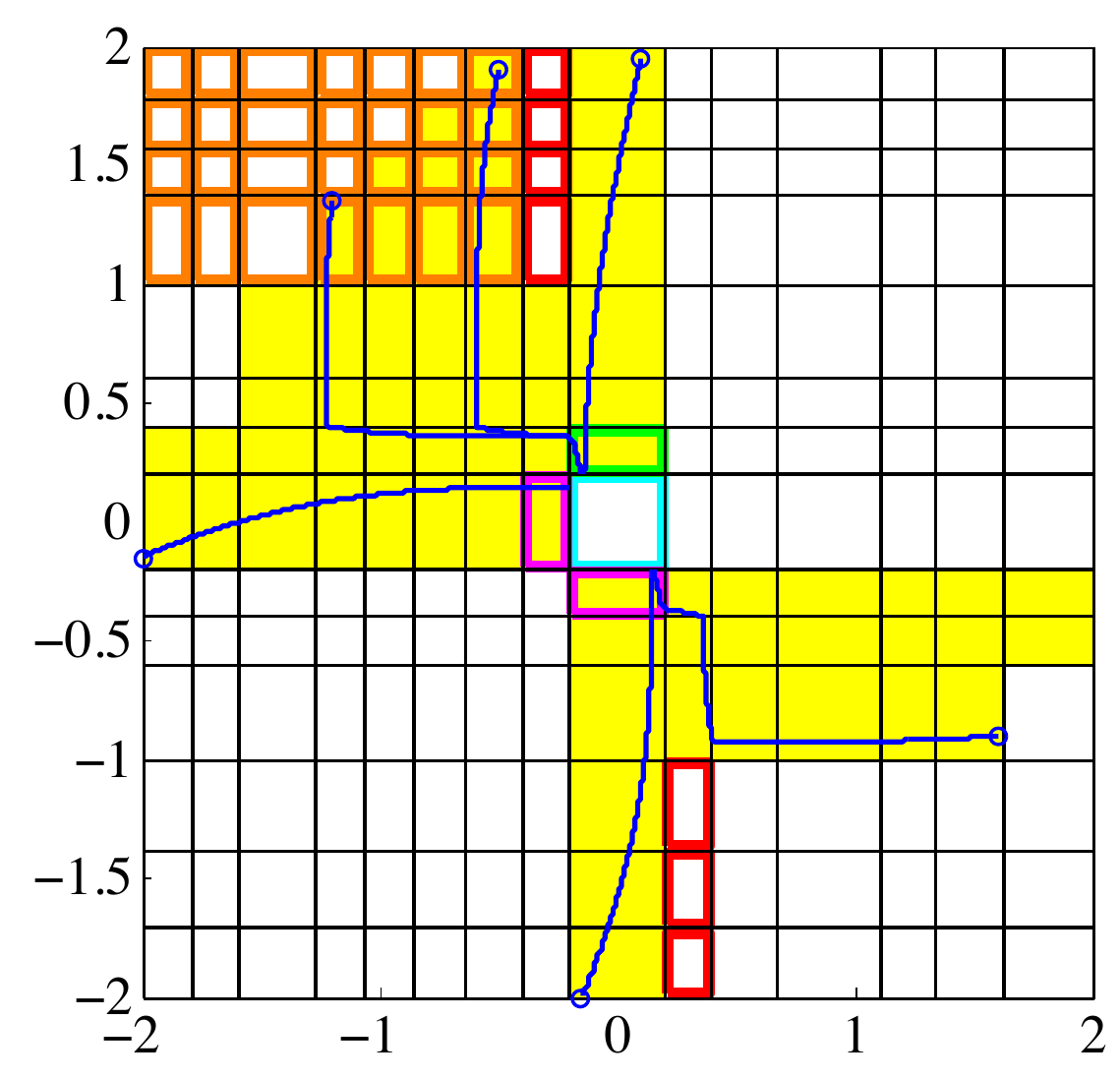}}
\subfloat[]{\label{fig:ex_nondet}\includegraphics[width=0.45\columnwidth]{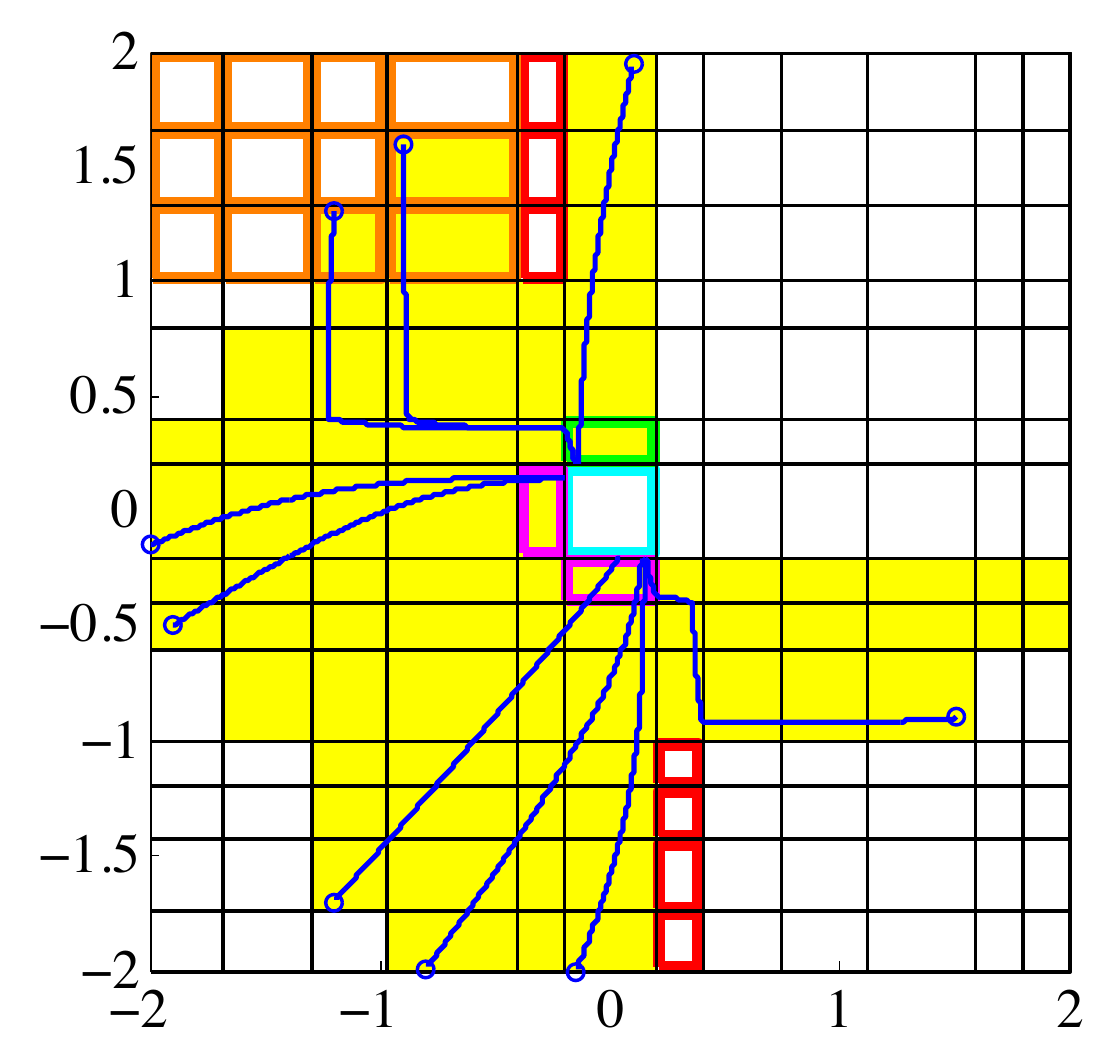}}
\caption{The yellow regions in (a) and (b) represent $\XO^d$ and $\XO^{nd}$, respectively. Some simulated satisfying trajectories of the corresponding closed-loop systems are shown (the initial states are marked by circles).}
\label{fig:sim}
\end{figure}

We use $(Q_0^d,\AUTOMATON^C_{d})$ and $(Q_0^{nd}, \AUTOMATON^C_{nd})$ to denote the control strategies as defined in Sec.~\ref{sec:controlStrategy} for the partition schemes resulted from the iterative refinement described in Sec.\ref{sec:optimization} for the first and second approach, respectively. We use $\XO^{d}$ and $\XO^{nd}$ to denote the corresponding sets of initial states of system\eqref{eq:case_system}, respectively. 
These sets, together with sample trajectories of the closed loop systems, are shown in Fig.~\ref{fig:sim}. 
The volume of $\XO^d$ is $5.25$ and the volume of $\XO^{nd}$ is $7.62$. 
A control-to-facet problem on a rectangle $R_2(a,b) \subseteq [-2,-0.2]\times [-2,-0.2]$ does not have a solution for facets $F^{e_1}$ and $F^{e_2}$ because of the strong drift in that region. However, rectangles in the same region have solutions to control-to-set-of-facets problem for $\Fset = \{F^{e_1}, F^{e_2}\}$. Consequently, rectangles in that region is only covered by $\XO^{nd}$ as the construction of $\XO^{nd}$ considers non-determinism.

\section{Conclusion}\label{sec:conclusion}

We studied a time-constrained control problem for a continuous-time multi-affine system from a specification given as a syntactically co-safe LTL formula over a set of predicates in its state variables. 
Our approach was based on finding an optimal control strategy on the product between an abstraction of the system and an automaton enforcing the satisfaction of the specification. The abstraction was a weighted transition system constructed by solving facet reachability problems on a rectangular partition of the state space of the original system. We proposed an iterative refinement procedure via a random optimization algorithm to increase the set of admissible initial states.

%\bibliography{multi-affine}

\begin{thebibliography}{21}
\providecommand{\natexlab}[1]{#1}
\providecommand{\url}[1]{\texttt{#1}}
\expandafter\ifx\csname urlstyle\endcsname\relax
  \providecommand{\doi}[1]{doi: #1}\else
  \providecommand{\doi}{doi: \begingroup \urlstyle{rm}\Url}\fi

\bibitem[Asarin and Maler(2009)]{Asarin:1999}
E.~Asarin and O.~Maler.
\newblock As soon as possible: Time optimal control for timed automata.
\newblock In \emph{Hybrid Systems: Computation and Control}, pages 19--30.
  Springer, 2009.

\bibitem[Belta(2004)]{Belta-ICRA2004}
C.~Belta.
\newblock On controlling aircraft and underwater vehicles.
\newblock In \emph{{IEEE} International Conference on Robotics and Automation},
  volume~5, pages 4905 -- 4910, 2004.

\bibitem[Belta and Habets(2006)]{Belta-TAC06}
C.~Belta and L.C.G.J.M. Habets.
\newblock Control of a class of nonlinear systems on rectangles.
\newblock \emph{IEEE Transactions on Automatic Control}, 51\penalty0
  (11):\penalty0 1749 --1759, 2006.

\bibitem[Bhatia et~al.(2010)Bhatia, Kavraki, and Vardi]{Kavraki:MPlanning}
A.~Bhatia, L.~E. Kavraki, and Moshe~Y. Vardi.
\newblock Motion planning with hybrid dynamics and temporal goals.
\newblock In \emph{{IEEE} Conference on Decision and Control}, pages
  1108--1115, 2010.

\bibitem[de~Jong(2002)]{jong2002}
H.~de~Jong.
\newblock Modeling and simulation of genetic regulatory systems.
\newblock \emph{J. Comput. Biol.}, 9\penalty0 (1):\penalty0 69--105, 2002.

\bibitem[Gazit et~al.(2007)Gazit, Fainekos, and Pappas]{Hadas-ICRA07}
H.~Kress Gazit, G.~Fainekos, and G.~J. Pappas.
\newblock Where's {W}aldo? {S}ensor-based temporal logic motion planning.
\newblock In \emph{{IEEE} International Conference on Robotics and Automation},
  2007.

\bibitem[Girard(2010{\natexlab{a}})]{Girard:2010}
A.~Girard.
\newblock Synthesis using approximately bisimilar abstractions: state-feedback
  controllers for safety specifications.
\newblock In \emph{Hybrid Systems: Computation and Control}, pages 111--120.
  ACM, 2010{\natexlab{a}}.

\bibitem[Girard(2010{\natexlab{b}})]{Girard:2010Opt}
A.~Girard.
\newblock Synthesis using approximately bisimilar abstractions: time-optimal
  control problems.
\newblock In \emph{{IEEE} Conference on Decision and Control}, pages 5893
  --5898, 2010{\natexlab{b}}.

\bibitem[Habets et~al.(2006)Habets, Kloetzer, and Belta]{Habets2006}
L.C.G.J.M. Habets, M.~Kloetzer, and C.~Belta.
\newblock Control of rectangular multi-affine hybrid systems.
\newblock In \emph{{IEEE} Conference on Decision and Control}, pages 2619
  --2624, 2006.

\bibitem[Kloetzer and Belta(2008{\natexlab{a}})]{Kloetzer:2008}
M.~Kloetzer and C.~Belta.
\newblock A fully automated framework for control of linear systems from
  temporal logic specifications.
\newblock \emph{IEEE Transactions on Automatic Control}, 53\penalty0
  (1):\penalty0 287 --297, 2008{\natexlab{a}}.

\bibitem[Kloetzer and Belta(2008{\natexlab{b}})]{Kloetzer:2008NonDet}
M.~Kloetzer and C.~Belta.
\newblock Dealing with nondeterminism in symbolic control.
\newblock In \emph{Hybrid Systems: Computation and Control}, pages 287--300.
  Springer-Verlag, 2008{\natexlab{b}}.

\bibitem[Kupferman and Vardi(2001)]{Vardi:safety}
O.~Kupferman and M.~Y. Vardi.
\newblock Model checking of safety properties.
\newblock \emph{Formal Methods in System Design}, 19:\penalty0 291--314, 2001.

\bibitem[Latvala(2003)]{Latvala:scheck}
T.~Latvala.
\newblock Efficient model checking of safety properties.
\newblock In \emph{In Model Checking Software. 10th International SPIN
  Workshop}, pages 74--88. Springer, 2003.

\bibitem[Lotka(1925)]{Lotka1925}
A.~Lotka.
\newblock \emph{Elements of physical biology}.
\newblock Dover Publications, Inc., New York, 1925.

\bibitem[Mazo and Tabuada(2011)]{Mazo:2011}
M.~Mazo and P.~Tabuada.
\newblock Symbolic approximate time-optimal control.
\newblock \emph{Systems and Control Letters}, 60\penalty0 (4):\penalty0 256 --
  263, 2011.

\bibitem[Nijmeijer and van~der Schaft(1990)]{vanderSchaft}
H.~Nijmeijer and A.J. van~der Schaft.
\newblock \emph{Nonlinear Dynamical Control Systems}.
\newblock Springer-Verlag, 1990.

\bibitem[Tabuada and Pappas(2003)]{TP03}
P.~Tabuada and G.~Pappas.
\newblock Model checking {LTL} over controllable linear systems is decidable.
\newblock In \emph{Lecture Notes in Computer Science}. Springer-Verlag, 2003.

\bibitem[Trelea(2003)]{PSOTralea}
I.~C. Trelea.
\newblock The particle swarm optimization algorithm: convergence analysis and
  parameter selection.
\newblock \emph{Information Processing Letters}, pages 317 -- 325, 2003.

\bibitem[Volterra(1926)]{Volterra1926}
V.~Volterra.
\newblock Fluctuations in the abundance of a species considered mathematically.
\newblock \emph{Nature}, 118:\penalty0 558--560, 1926.

\bibitem[Wolfgang(2002)]{Wolfgang2002}
T.~Wolfgang.
\newblock Infinite games and verification.
\newblock In \emph{Computer Aided Verification}, pages 58--65. Springer Berlin
  / Heidelberg, 2002.

\bibitem[Wongpiromsarn et~al.(2009)Wongpiromsarn, Topcu, and
  Murray]{Tok-Ufuk-Murray-CDC09}
T.~Wongpiromsarn, U.~Topcu, and R.~M. Murray.
\newblock Receding horizon temporal logic planning for dynamical systems.
\newblock In \emph{{IEEE} Conference on Decision and Control}, pages
  5997--6004, 2009.

\end{thebibliography}

\end{document}